\documentstyle[12pt,epsf]{article}
\def \ins#1#2#3#4#5#6#7 {
  \vspace{-#4}
  \epsfxsize #3
  \epsfysize #4
  \begin{figure}[#6]
    \centerline{\epsfbox{#2}}
    \caption{#7}
    \label{#1}
  \end{figure}}

\begin{document}
\begin{center}
{\Large \bf Simulation of Nucleus-Nucleus Interactions
in the Framework of the FRITIOF Model at the Energy of 3.3 GeV/nucleon}
\end{center}

\begin{center}
A.S. Galoyan
\footnote{
Yerevan Physical Institute and
Joint Institute for Nuclear Research, e-mail: galoyan@cv.jinr.ru},
A. Polanski
\footnote{Soltan Institute for Nuclear Studies, 05-400 Swierk,
Poland, e-mail: polanski@cv.jinr.ru},
V.V. Uzhinskii
\end{center}

\begin{center}
Joint Institute for Nuclear Research, \\
Laboratory of Information Technologies
\end{center}

\begin{center} Abstract \end{center}

\begin{center}
\begin{minipage}[c]{12cm}{\small
The intranuclear cascade model overestimates the
multiplicity of produced mesons in nucleus-nucleus interactions
without taking into account meson and baryon resonance production.
Inclusion of the resonances leads to decreasing multiplicity of
mesons, neutrons and protons. In order to overcome the problem, it is
proposed to use the FRITIOF model adapted to low energies in a
combination with the reggeon theory inspired model of nuclear
destruction. It is shown that the combination allows one to reproduce
satisfactory the meson and baryon yields in the nucleus-nucleus
collisions at the energy of 3.3 GeV/nucleon. The combined model works
faster then typical quantum molecular dynamic model, and allows one to
estimate the data needed for creation of electro-nuclear amplifier.}
\end{minipage}
\end{center}

\section{ Introduction}

Data on nuclear reactions at the energies of hundred MeV and GeV are
required for multiple purposes such as long-lived radioactive waste
transmutation, material analysis, nuclear medicine as well as research
of the cosmic ray effects on spaceships and astronauts. Experiments on
measuring the data are costly to be carried out and there is a
limited number of facilities to make them. Therefore, reliable
computer models for a simulation of the reactions are created to
provide the necessary data.  Most of them are using the ideas of the
cascade-evaporation model (CEM) (see \cite{CEM_1} - \cite{CEM_6}).

The cascading of nucleons and $\pi$-mesons was considered only by the
first variants of the model. A good description of hadron-nucleus
interactions was reached at the studies. The best result was obtained
by S. Mashnik \cite{Mashnik94}. Though, the application of the model for
description of nucleus-nucleus collisions has shown that the model
gives a satisfactory yield of the nucleon but overestimates the meson
production. Taking into account meson and baryon resonances production
is one of the possible ways to solve the problem.

A number of authors were trying to do this, and they usually obtained
a decreasing yield of mesons and baryons. It is natural
because the effective decreasing of the multiplicity of the produced
particles leads to a less powerful cascading. So, a problem of a self
consistent description of the meson and baryon yields in hadron-nucleus
and nucleus-nucleus interactions appeared.

Let us note that the nuclear destruction mechanism and the procedure of
the excitation energy calculation were not changed in the mentioned
approaches. Maybe, they led to the unsatisfactory results. In this
paper we consider a synthesis of the FRITIOF model
\cite{Fritiof1, Fritiof2} which takes into account the resonances
production and the reggeon theory inspired model of the nuclear
destruction \cite{RTIM, Olek}.

The FRITIOF code \cite{Fritiof2}, that is a program of
Monte Carlo simulation of the inelastic hadron-hadron, hadron-nucleus
and nucleus-nucleus interactions, is very popular in high energy
experimental physics. It is explained by its access,
its physical ideas simplicity and a defined beauty of the code by
itself. It is easy to  use.

The FRITIOF model \cite{Fritiof1, Fritiof2} assumes that an excitation
of hadrons into continuum mass spectra takes place in the inelastic
hadron-hadron collisions. In the case of the hadron-nucleus or the
nucleus-nucleus interactions the excited hadrons can suffer an
additional collisions with the nuclear nucleons and go into more
excited states, or de-excite. The excited hadrons are considered as
quark strings, and the corresponding quark model \cite{JETSET2,
JETSET3} is used for description of their decay. The probabilities of
the multiple collisions are calculated within the
Glauber model \cite{Glauber59} -- \cite{DIAGEN}.
The inelastic collisions are usually considered only \cite{DIAGEN}. In
order to reproduce the baryon yield, we introduce elastic re-scattering
too.

It is assumed that the program can not be used at the relatively low
energies as the hypothesis about the creation and decay of the quark
strings is not valid. Attempts of compelling the program to operate at
the energies below 5-10 GeV/nucleon for AA-interactions usually failed.
Though the analysis of the code operation shows that the
program cycles due to its quite simple-hearted interpretation of the
Fermi-motion of the nucleons. A change of the Fermi-motion simulation
algorithm \cite{Adamovich} allowed one to decrease a formal limit of the
model application region.

Now in order to use the model at the intermediate energies, one needs
to correct a scheme of the fragmentation of the quark string with low
masses because in this case the excited hadrons have low masses, too.
Below the corresponding changes of the model will be presented which
allow one to reach a defined success. To solve the problem, we have
used the experimental data on the neutron-proton interactions at
momentum 1.25 -- 5.1 GeV/c \cite{Troyan} given amiable by the Yu. A.
Troyan's group of the neutron-proton interaction study.

One of the disadvantages of the FRITIOF code is an omission of the slow
particle cascading into the nuclei. Under the "cascading" one usually
understands the standard intranuclear cascade scenario
(see, e.g., \cite{CEM_1}), which neglects the quantum mechanical
effects. The quantum mechanical description of the particles cascading
in the nuclei can be achieved in the framework of the reggeon
theory. According to the reggeon approach \cite{BKKS}, a consideration
of the cascade interactions necessitates a calculation of the yields of
the so-called enhanced diagrams to the elastic scattering amplitude.
Using the procedure for the calculation proposed in \cite{KPTM} and
Abramovski-Gribov-Kancheli cutting rules \cite{AGK}, one can
obtain the positive defined cross-sections of the inelastic processes.
At first glance, this returns one to the classical cascade picture of
the interactions. However, there is an essential difference.  The
cascade model assumes that the cascade is developed in
a three-dimensional space of a target nucleus. According to the reggeon
approach \cite{RTIM}, the "cascade" of the reggeon exchanges occurs in
a two-dimensional space of projected radius-vectors of nucleons on the
plane perpendicular to the momentum of projectile particle (on a plane
of impact parameter), with a "cascade power" independent of the
multiplicity of the produced particles and defined by the reggeon
vertex constants and the size of nucleus. We shell give a
corresponding algorithm in Sec. 3.  The main calculation results are
presented in Sec. 4. We are starting with a short description of the
main assumptions of the FRITIOF model.

\section{Theses of the FRITIOF model}

The FRITIOF model assumes that the hadron-hadron interactions are of
two-particle character
\begin{equation}
a + b \longrightarrow a'+ b', \label{1}
\end{equation}
where $a'$ and $b'$ are $a$ and $b$ hadrons in excited states.
The kinematics of the reaction is determined as follows:
in the center of mass of colliding hadrons the
energy-momentum conservation low has the form:
\begin{eqnarray}
E_a + E_b & = & E_{a'} + E_{b'} = \sqrt{s_{ab}}, \nonumber \\
p_{a z} + p_{b z} & = & p_{a' z} + p_{b' z} =0, \label{2}\\
0 & = & \vec p _{a' \perp} + \vec p _{b' \perp}, \nonumber
\end{eqnarray}
where $E_a$ and $E_b\  (E_{a'},E_{b'}) $- energies of the initial
(final) hadron $a$ and $b\ (a',b')$, $p_{a z}$ and $p_{b z}$ -
longitudinal momentum components (projected momenta on the interaction
axis).

Adding and subtracting the first
two equations from (\ref{2}), we get
\begin{eqnarray}
P^+ _a + P^+ _b &=& P^+ _{a'} + P^+ _{b'} \nonumber \\
P^- _a + P^- _b &=& P^- _{a'} + P^- _{b'} \label{3} \\
0 & = & \vec p _{a' \perp} + \vec p _{b' \perp}, \nonumber
\end{eqnarray}
where $ P^+ = E + p_z ,~~~ P^- = E -p_z.$

At high energies:
$
P^- _{a'} \simeq m^2 _{a'} / 2 \mid p_{a' z} \mid ,~~
P^+ _{b'} \simeq m^2 _{b'} / 2 \mid p_{b' z} \mid .$
Thus, the $P^-_{a'}$ and $P^+_{b'}$ distributions used in the code have
the form
\begin{eqnarray}
dW & \sim & dP^- _{a'}
/P^- _{a'} \simeq d m^2 _{a'} /m^2 _{a'} , \nonumber \\
dW & \sim & dP^+ _{b'} /P^+ _{b'} \simeq d m^2 _{b'} /m^2 _{b'} .
\label{5}
\end{eqnarray}
The limits of $P^-_{a'}$ and $P^+_{b'}$ are defined as
\begin{equation}
[ P^- _{a}, P^- _{b}],~~~ [ P^+ _{b}, P^+ _{a}]. \label{6}
\end{equation}
The distributions (\ref{5}) are typical for the so-called high-mass
diffraction dissociation processes.

Knowing $P^-_{a'}$, $P^+_{b'}$, $\vec p _{a' \perp}$, $\vec p _{b'
\perp}$ and determining  $P^+_{a'}$, $P^-_{b'}$ from Eqs. (\ref{3} ),
one can find the masses of the excited hadrons $a'$ and $b'$.  $\vec p
_{a' \perp}$ and $\vec p _{b' \perp}$ are sampled according to the low
\begin{equation}
dW  \sim  exp(-\vec p^2 _{a' \perp}/<p^2_{\perp}>)~d^2 p _{a' \perp}.
\label{7}
\end{equation}

In case of hadron-nucleons interactions the kinematics governed by
Eqs. (\ref{3}), (\ref{5}), (\ref{7}) is applied to the first collision
of the projectile nucleon with one of the target nucleons ($a + N_1
\rightarrow a' + N'_1$). For the second collision ($a' + N_2
\rightarrow a'' +N'_2$), analogous relations are used, but (\ref{6}) is
replaced for
\begin{equation}
[ P^- _{a'}, P^-_{N_2}],~~~ [ P^+ _{N_2}, P^+ _{a'}].
\label{8}
\end{equation}
As a result, the consequent collisions involve
a systematic increasing of the mass of hadron $a$ if transfers of
the transverse momentum are small.

A similar approach is also applied to simulate the nucleus-nucleus
interactions. Here the reactions $a' + b' \rightarrow a''+ b''$
are considered. The above distributions on $P^-_{a'}$ and $P^+_{b'}$
are replaced by those on $P^-_{a''}$ and $P^+_{b''}$ and the limits of
$P^-_{a''}$ and $P^+_{b''}$ are redefined as
\begin{equation}
[ P^- _{a'}, P^- _{b'}],~~~ [ P^+ _{b'}, P^+ _{a'}].
\label{9}
\end{equation}

At relatively lower energies of the order of 5--10 GeV/nucleon,
the FRITIOF model without taking into account hadron de-excitation
overestimates the multiplicity of the produced particles in hA- and
AA-interactions. To this end we have changed the condition (\ref{9})
by the following giving an allowance for considering the excitation
process with the increasing mass and the de-excitation process with
decreasing mass:
\begin{equation}
[ P^- _{a}, \sqrt{s_{a'b'}} - m_b],~~~ [ P^+ _{b},
\sqrt{s_{a'b'}}-m_a].
\label{10}
\end{equation}
$P^-_a$ and $P^+_b$ are calculated at
$s_{ab} = s_{a'b'},~~ m_{a'}=m_a,~~ m_{b'}=m_b$. The minimal values of
$P^-_a$ and $P^+_b$, being the lower limits in (\ref{10}), are
obviously reached in the reaction $a' + b'\rightarrow a + b$.  The
maximal values are achieved at $p_{a''z} =p_{b''z} = 0$ when hadrons
come to rest in the c. m. frame of $NN$-collision.

The reactions (\ref{1}), or
\begin{eqnarray}
a + b \longrightarrow a'+ b, \label{pdif}\\
a + b \longrightarrow a + b', \label{tdif}
\end{eqnarray}
are the so-called diffraction dissociation reactions. The reactions
(\ref{pdif}), (\ref{tdif}) are one-vertex diffractions, the reaction
(\ref{1}) is a double vertex diffraction. It is obvious that a minimal
mass of the excited nucleon in NN-interactions can be equal to
$m_{a'}=m_N + N_{\pi} = 1080$ MeV. In the FRITIOF model the minimal
mass is equal to 1.2 GeV. Due to this, the one vertex diffraction with
excitation only one hadron is possible at
$\sqrt {s_{NN}} \leq 2.4$~GeV, or at $P_{lab} \leq 1.91$ GeV/c
in NN-interactions. There can be a two vertex diffraction at higher
energies. A relation between the cross-sections of the processes is
determined by the hadrons $a^\prime $ and $b^\prime$ mass
distributions.

The excited hadrons $a^\prime$ and $b^\prime$ are considered as
quark strings, and the corresponding quark model is used for the
simulation of their decay \cite{JETSET2, JETSET3}. It is assumed that
the quark model can be used at sufficient large string masses what can
be created at high energies. Thus the FRITIOF model was used mainly at
high energies. One can expect that the FRITIOF model predictions fall
into a contradiction with experimental data on the processes where the
states with low masses can appear. In order to study the situation,
we turn to the data on $\pi ^-$-meson and proton distributions in
$np$-interactions at $P_n=$ 1.25--5.1 GeV/c \cite{Troyan}.

Fig. \ref{fig1} shows the experimental and calculated $\pi ^-$-meson
rapidity distributions, $y= \frac{1}{2}\ln (E+P_z)/(E-P_z),$
where $E$ and $P_z$ are laboratory energy and longitudinal momentum of
$\pi ^-$-meson, respectively. As seen, the distributions calculated
according to the original code (dashed lines) are of two-bump structure
more pronounced at low energies. It seems like it is a circumstance of
the assumed diffractive character of the interactions: the bump at
large rapidities is caused by a projectile hadron diffraction, the
bump at low rapidities is connected with a target nucleon diffraction.
Though, at the neutron momentum of 1.25 GeV/c a diffraction system with
mass of 1.2 GeV (the minimal mass of excited nucleon assumed by the
FRITIOF model) must be in a rest in the center of mass system, and
there must not be a subdivision of the fragmentation regions. Thus,
we conclude that the two-bump structure of the calculated distributions
is not a circumstance of the diffraction dissociation of the hadrons.
It only reflects an anysotropy of the low mass string decay. A direct
simulation of the decay of the strings with low masses has shown that
the bi-module structure is typical for the decay of the strings with
masses lower than 1.7 GeV. Two-particle channel is dominating in the
decay of such strings. At higher masses the multi-particle channel
gets more probable, and the calculated distributions get more
regular.

Taking into account the character of the experimental distributions at
$P_n$ = 1.25, 1.73 GeV/c, it seems reasonable to simulate an isotropic
decay of the strings with low masses in the case of the two-particle
decay channel. A boundary value of the string mass of 1.7 GeV  below of
which we have used the proposed procedure, was  chosen requiring a
good description of the data.  As seen in the Fig. \ref{fig1} (solid
curves), this allows us to describe the experimental data quite well.

A more complete situation appears in the description of the $\pi
^-$-meson distributions in transverse momentum, $P_T$. The original
model predicted the average transverse momentum larger than the
experimental ones (see the dashed curves in Fig. \ref{fig2}), though we
had changed the character of the low mass string decay. Since at $P_n
= 1.25$, we simulate an isotropic decay of the excited hadrons, the
maximum of the distribution in $P_T$ is determined by mass of the
decayed system. Thus, in order to decrease the average transverse
momentum, the minimal mass of the excited nucleon state was decreased
to the value of 1.1 GeV. This, as seen from Fig. \ref{fig2}, gave a
better result at $P_n < 2$ GeV/c. At larger energies one needs to
take into account the transverse momentum transferred by the colliding
nucleons.

An analysis of proton spectra gives some additional information, and
allows one to determinate the model parameters more exactly. For the
analysis we have used the data on the following reactions
\begin{eqnarray}
np & \longrightarrow & pp\pi^-;      \label{11} \\
np & \longrightarrow & pp\pi^-\pi^0; \label{12} \\
np & \longrightarrow & np\pi^+\pi^-. \label{13}
\end{eqnarray}

In the reaction (\ref{11}) according to the model the diffraction
dissociation processes of the projectile particle is a dominant one.
In the reaction (\ref{12}) the two-vertex diffraction processes do
the same. At last, in the reaction (\ref{13}) we have equal yields of
the one-vertex and two-vertex diffraction. Thus, the study of the
reactions allows one to check the different components of the model.

Fig. \ref{fig3} gives the experimental and calculated rapidity
distributions of the protons. The experimental data are
presented by the histograms, the original model calculations - by the
dashed curves, and the last calculations - by the solid curves. As
seen, there is a two-bump structure of the experimental distributions
on the reactions (\ref{11}). The bump at small rapidities is caused by
the saved target protons. The bump at large rapidities is connected
with the protons created in the projectile particle diffraction. The
distribution of the reaction (\ref{12}) has no structure. At last, in
the reactions (\ref{13}) there is a dominant production of the protons
in the target fragmentation region.

The original model calculations are in agreement with the data on the
reaction (\ref{13}). At the same time, the calculated distribution for
the reaction (\ref{11}) has a bump in the region $y \sim 1.7$ what does
not observe at the experiment. For its elimination we introduce a
charge exchange between the nucleons in 50 \% of the two-vertex
diffraction. This allowed us to improve the proton spectra description
in part.

\underline{Enumeration of the changes made in the FRITIOF code}

\begin{enumerate}

\item The minimal mass of the excited nucleon decreases from 1.2 GeV
to 1.1 GeV;

\item In case of the two-particle decay channel of a string with a
mass lower than 1.7 GeV the isotropic decay is simulated in the center
of the mass system;

\item The charge exchange between the colliding nucleons is allowed in
50~\% of the two-vertex diffraction;

\item
The value of the average square of the transverse momentum which
transferred between the colliding nucleons increases from 0.08
(GeV/c)${}^2$ to 0.15  (GeV/c)${}^2$.

\end{enumerate}

\section{Simulation of nuclear destruction at the fast stage
of interaction.}
\subsection{Determination of the number of knocked-out nucleons}

In the last few years there have been some successful attempts to
describe the hadron-hadron elastic scattering at low and intermediate
energies (below $1 - 2~ GeV$) within the quark-gluon
approach (see Refs. \cite{Barnes1} - \cite{Barnes4}). In Ref.
\cite{Barnes1} -\cite{Barnes4} the amplitudes of $\pi  \pi -$, $K \pi
-$ and $NN-$ scattering were found and an agreement of the theoretical
calculations with corresponding experimental data was reached at the
assumption that in the elastic hadron scattering one-gluon exchange
with the following quark interchange between hadrons takes place (see
Fig. \ref{fig4}a).  At high energies two-gluon exchange appropriation
(Fig.\ref{fig4}b) works quite well (see Ref.  \cite{Low},
\cite{Nussinov} and \cite{Gunion}).  What kind of
exchanges can dominate in hadron-nucleus and nucleus-nucleus
interactions?

The simplest possible diagrams of the processes with three
nucleons are given in Fig. \ref{fig5}. Calculation of their amplitudes
according to Refs. \cite{Barnes1}-\cite{Barnes4} is a serious
mathematical problem. It can be simplified if one takes into account an
analogy between the quark-gluon diagrams and the reggeon diagrams: the
quark diagram of Fig. \ref{fig4}a corresponds to a
one-nonvacuum-reggeon exchange diagram; the diagram of Fig. \ref{fig4}b
describes the pomeron exchange in the $t-$ channel; the diagram of Fig.
\ref{fig5}a is in a correspondence with the enhanced reggeon diagram of
the pomeron splitting into two non-vacuum reggeons.  The three pomeron
diagram (Fig. \ref{fig5}d) represents a more complicated process. It
is rather hard to find a correspondence between the reggeon diagrams
and the diagrams of Fig.  \ref{fig5}b, \ref{fig5}c.

The reggeon parameters and the functional forms of the amplitudes of
3-reggeon processes are well known. The constants of the reggeon
interaction vertexes are poor determined. The 3-pomeron vertex
constant $G_{PPP}$ is well established ($G_{PPP}=1.35^{-2} (GeV)^2$,
Ref.  \cite{KPTM}). There are only old data \cite{Kasar} and the
estimations of Ref. \cite{Volkov} on the values of other constants -
$G_{PRR}$ and
$G_{RRR}$, which are large. Nevertheless, we believe that the properties
of the reggeon amplitudes must be taken into account in
consideration of the nuclear destruction.

It is obvious that the processes like that in Fig. \ref{fig5}d can not
dominate in the elastic hadron-nucleus scattering because they are
accompanied by production of a high mass diffraction beam of the
particles in the intermediate state. Thus, their yields are dumped by a
nuclear form-factor. According to the same reason, the yields of the
processes like ones in Figs. \ref{fig5}a,  \ref{fig5}b can be small,
too. If it is not so, one will expect large corrections to Glauber's
cross-sections. The practice shows that the corrections to the
hadron-nucleus cross-sections must be lower than 5~--~7~$\%$.

The yield of the diagram of Fig.~\ref{fig5}c gives a correction to
Glauber's one-scattering amplitude. There must be analogous
corrections to the other terms of Glauber's series. The sum of the
corrections must lead to small effects in the elastic small angle
scattering because the corrections are large at small impact
parameters. So, they can manifest themselves at large scattering
angles. We assume that they have a big influence on the inelastic
process characteristics, too.

According to the reggeon theory, a description of the inelastic
reactions can be reached in a consideration of the different cuts of
the reggeon diagrams. Here the Abramovski - Gribov - Kancheli cutting
rules \cite{AGK} are often used. The corrections to them were
discussed in Ref. \cite{BKKS} in the application to the problem of
a particle cascading on the nucleus. As was shown in Ref.
\cite{BKKS}, summation of the yields of enhanced diagrams allows
one to describe increasing of the one-particle spectra in the target
fragmentation region. At the same time, the authors of Ref.
\cite{BKKS} did not take into account the shadowing effects
considered in Ref. \cite{Shad}.

Here we have to note that the yields of the diagrams like that shown in
Fig.~\ref{fig5}c have no shadowing corrections. The yield of the
enhanced diagram of Fig. \ref{fig5}a has a form
$$
Y_a \sim exp[ -(\vec b _1 - \vec b _2 )^2 /3 r_a ^2 -
(\vec b _1 - \vec b _3 )^2 /3 r_a ^2 - (\vec b _2 - \vec b _3 )^2 /3
r_a ^2 ]
$$
where $\vec b _1 , \vec b _2$ and $\vec b _3$ are the impact
coordinates of the nucleons. At the same time, the yield of the diagram
of Fig. \ref{fig5}c according to Refs. \cite{Barnes1} - \cite{Barnes4}
is given by
$$
Y_c \sim exp[ -(\vec b _1 - \vec b _2 )^2 / r_c ^2 ]
exp[ -(\vec b _2 - \vec b _3 )^2 / r_c ^2 ].
$$
In the limit of $r_a^2 , r_c ^2 \ll R_A ^2$, where $R_A$ is a nucleus
radius, the yields coincide.  Thus, we can save the results of Ref.
\cite{BKKS} considering them as a summation of the yields of the
quark-gluon diagrams.

Let us note that neither $Y_a$, nor $Y_c$ depend on the
longitudinal coordinates or on the multiplicity of produced particles.
It is the main difference between the ''reggeon cascading'' and
the ''usual'' cascading.

As well known, the intranuclear cascade model
(\cite{CEM_1}-\cite{CEM_6}) assumes that in a hadron-nucleus collision
the secondary particles are produced due to an inelastic interaction of
the projectile particle with a target nucleon. The produced particles
can interact with other target nucleons. A distribution on distance
$l$ between the first interaction and the second one has a form
$$
W(l)~dl \sim \frac{n}{<l>} exp(- \frac{n}{<l>} l),
$$
where $<l>=1/\sigma \rho _A$, $\sigma$ is a hadron-nucleon
cross-section, $n$ is the multiplicity of the produced particles and
$\rho _A \simeq 0.15 fm^{-3}$ is the nuclear density. At the same time
the amplitudes or the cross-sections of the processes shown in
Fig.\ref{fig5} have no dependence on $l$ or $n$. Thus, we expect that
in the quark-gluon or reggeon approach the ''cascade'' will be more
restricted than in the cascade model. The difference between
approaches can lead to the different predictions for the light nuclei
destruction (an effect of the limited volume) and for the
characteristics of the heavy nuclei interactions (an influence of a
large multiplicity of the produced particles).

To show this, we use a simple method to estimate
the nuclear destruction in the framework of the quark-gluon approach.
\begin{enumerate}
\item
As it was said above, the ''reggeon cascade'' is developed in the space
of the impact parameter. Thus, for its description it is needed to
determinate a probability to involve a nucleon into the ''cascade''. It
is obvious that the probability depends on a difference of the impact
coordinates of the new and the previously involved nucleons. Looking at
the yield of the diagram of Fig. \ref{fig5}c, we choose the functional
form of the probability as
\begin{equation}
P(\mid \vec b _i - \vec b _j \mid ) = C_{nd} exp( - (\vec b _i -
\vec b _j )^2 /r_nd ^2 ).
\label{pij}
\end{equation}
Here $\vec b _i$ and $\vec b _j$ are projections of the radiuses of
$i^{th}$ and $j^{th}$ nucleons on the impact parameter plane.

\item The ''cascade'' is initiated by the primary involved, wounded
nucleons.  If the constant $C_{nd}$ is small, we can use the Glauber
theory for their determination.

\item We assume that all the involved and wounded nucleons are ejected
from the nucleus.
\end{enumerate}

The ''cascade'' looks as follows: a projectile particle interacts with
some of the intranuclear nucleons. They are called ''wounded''
nucleons. The wounded nucleons initiate the ''cascade''. A wounded
nucleon can involve a spectator nucleon into the ''cascade'' with the
probability (\ref{pij}). The latter can involve a second nucleon.
The second nucleon can involve a third one, and so on.

A Monte Carlo algorithm for estimation of the nuclear destruction in
the nucleus-nucleus interactions which corresponds to the model
formulation, includes the following steps:
\begin{enumerate}
\item The calculation of the impact parameter distribution in the
framework of the Glauber theory \cite{DIAGEN};
\item The sampling of the impact parameter and the nucleon coordinates;
\item The determination of the wounded nucleons (see Ref.  \cite{DIAGEN});
\item The determination of the spectator nucleons involved in the ''cascade''
by the wounded nucleons. If the number of the involved nucleons is
equal to zero - exit;
\item If the number of the involved nucleons is not equal to zero,
a possibility is considered to involve the other spectators nucleons by
the involved ones. If the number of the new involved nucleons is
equal to zero - exit. In other case - it is needed to repeat the step 5
taking into account only the new involved nucleons.
\end{enumerate}

The first step is performed only once at the given mass numbers of the
projectile and target nuclei. The steps 2 -- 5 are repeated until the
needed statistics is reached. The steps 4, 5 are applied to the
nucleons of projectile and target nuclei.

It is suggested that all the newly involved participants and
the "wounded" nucleons are knocked-out from the nucleus.

\subsection{Fermi-motion of nucleons}

To take into account the energy-momentum conservation law by
simulating compound system, let us consider a reaction of the compound
system $(1,2)$ with hadron $h$:  $(1,2) + h \rightarrow 1 + 2 +h$.
Neglecting the transverse momenta, a final state of the
reaction will be fully characterized by a value of merely one
independent kinematical variable. As the variable, let us take
$$
x^+ _1 = (E_1 + p_1)/(E_1 + E_2 + p_1 + p_2).
$$
It is useful to introduce the analogous quantity $x^+_2$
which satisfies obviously the relation $x^+_1+x^+_2=1$.  At given
$x^+_1$ and $x^+_2$, the rest of the kinematical variables can be
determined by the energy-momentum conservation law.

In case of dissociation of two compound systems $A$ and $B$ containing
$A$ and $B$ constituents respectively, let us introduce for the $i$-th
constituent of system $A$
$$
x^+ _i =(E_{Ai} + p_{iz})/W^+ _A~~~ and~~~\vec p _{i\perp},
$$
and for the $j$-th constituent of system $B$
$$
y^- _j =(E_{Bj} - q_{jz})/W^- _B~~~ and~~~\vec q _{i\perp},
$$
where, $E_{A_i} (E_{B_i})$ and $\vec p_i (\vec q_i )$ are energy and
momentum of $i$-th constituent from $A~(B)$,
$$
W^+ _A = \sum ^A _{i=1} (E_{Ai} + p_{iz}),~~~
W^- _B = \sum ^B _{i=1} (E_{Bi} - q_{iz}).
$$

One can find $W^+_A$ and $W^-_B$ using the energy-momentum conservation
low at given $\{x^+ _i,~\vec p _{i\perp}\}$, $\{y^- _i,~\vec q
_{i\perp}\}$, and determine all kinematical variables for all
constituents \cite{Adamovich}.

According to the experimental evidence \cite{PTfrag}, the average
transverse momentum of spectator fragments obeys the parabolic law:
$$
<P_{\perp} ^2 >=\frac {A (A-F)}{A} <p_{\perp} ^2 >,~~~ \sqrt{<p_{\perp}
^2 >} = 0.07~~ GeV/c.
$$

To reproduce this result, the values of $\vec p _{i\perp}$ for
knocked-out nucleons are simulated according to the distribution
\begin{equation}
dW \propto exp(- \vec p _{i\perp} ^2 /
<p_{\perp} ^2 >) d^2 p_{i\perp}, \sqrt{<p_{\perp} ^2 >}=0.07.
\label{pti}
\end{equation}

The sum of the transverse momenta (with sign "minus") was ascribed for
the residual nucleus.

The choice of $x^+_i$ is carried out by
\begin{equation}
dW \propto exp[- (x^+ _i -1/A)^2/(d_x /A)^2 ] d x^+ _i ,~~~
0\leq x^+ _i < 1,~~~ d_x =0.07.
\label{xi}
\end{equation}
$x^+$ of the residual
nucleus is determined as $1 - \sum  x^+ _i $ .

It was assumed that the knocked-out nucleons changed their
characteristics again. The new values of $x^+_i$ and $p_{i\perp}$ were
simulated using the distributions (\ref{pti}) and (\ref{xi}) at
$<p_{\perp} ^2 > = 0.385$ $(GeV/c)^2$ and $d_x=0.2$.  The results from
\cite{Propan1, Propan2, ABaldin} were used for determination of the
fitting parameters.

\subsection{Excitation energy of residual nucleus.}
The change of the nuclear destruction mechanism requires a change of
the nuclear residual excitation energy calculation procedure. In a
self-consistent diagram approach one needs to consider the more
complete diagrams and describes bound states of the nucleons. The last
problem is not solved in the presented quark-gluon approach which
allows one only to calculate an repulsive part of the NN-potential. It
is expected that a taking into account the terms with high order vertex
constant allows one to calculate the attractive part of the potential
too. Until this is not made, we are to use a phenomenological approach.
Here we follow the Ref. \cite{AbulMagd}.

In paper \cite{AbulMagd} proton-nucleus interactions at intermediate
energies were analyzed. The first stage of the interactions
was considered within the Glauber approach. It was assumed
that the projectile hadron undergoes successive interactions (elastic
and inelastic) with target nucleon. In each of the collisions a part
of the energy of the projectile hadron, $E$, is transferred to the
target nucleon. The distribution on the transfer energy was chosen in
the form:
\begin{equation}
F_{1}(E)=\frac{1}{\langle E \rangle}e^{-E/\langle E \rangle}.
\label{Erecol}
\end{equation}

The authors of Ref. \cite{AbulMagd} supposed that the excitation
energy of residual nucleus was the sum of the recoil nucleon energies.
As a result, they described experimental data on neutron multiplicity
dependence upon the excitation energy of the residual nuclei.

To apply this approximation to $AA$-interactions at high energies, one
needs to evaluate the number of re-scatterings of each of the
knocked-out nucleons. Taking into account that most of
$AA$-interactions are of peripheral nature, we assume that the slow
nucleons knocked-out from the peripheral parts of nucleus can not
penetrate deeply inside the nucleus because of the large $NN$
cross-section. Their re-scatterings thus occur in the nearest
environment. We consider the environment as the spectator nucleons
being inside a sphere with radius $r_0 = 2$ fm surrounding a wounded
or involved nucleon in its initial state. It is assumed that each
spectator nucleon may acquire recoil energy distributed according to
Eq. (\ref{Erecol}). Remember, that the nucleon coordinates were chosen
randomly and independent according to the Saxon-Woods distribution. If
a spectator nucleon was a neighbor of two "wounded" ones, or involved
nucleons, it received the sum of two energies chosen according to Eq.
(\ref{Erecol}), and so on. The sum of energies transferred to all
spectator nucleons was considered as the excitation energy of the
residual nucleus.

Unlike the CEM, this method obviously will lead to the
zero-excitation energy when all nucleons are be ejected.

It also seems evident that the boundary between the spectator part of
the nucleus and the part affected by the first stage of the interaction
which governs the excitation energy of the residual nucleus, depends on
the impact parameter. For example, with heavy projectile nuclei the
excitation energy rises with decreasing impact parameter from $R_A+R_B$
to 0, approaching a maximum and then falls. According to the CEM, it
must gradually increase.

We have used $\langle E \rangle =8$ MeV and the standard evaporation
model \cite{Weis37} (see, also Ref. \cite{CEM_1}) for simulation of the
residual nucleus de-excitation.

\section{Description of nucleus-nucleus interactions}
Fig. \ref{fig6} shows the proton and $\pi ^-$-meson rapidity
distributions in different nucleus-nucleus interactions.
They were calculated by the modified FRITIOF code and by the code of
the cascade-evaporation model\footnote{ ${}^)$The model takes into
account the trailing effect, Pauli principle, the dependence of the
Fermi momentum on the local nuclear density, the pre-equilibrium
emission and the evaporation of the nuclei.} \cite{Zhenis}. Events with
at least one inelastic NN-collision were selected at the simulations.
As seen, for $dd$-interactions the models give close results. Since
in this case we can neglect the cascade interactions, the coincidence
of the results tells us about correctness of the NN-interactions
description in the CEM
(the CEM code used by us does not allow a direct simulation of
NN-interactions). We can mark only a little enhanced baryon production
in the central rapidity region in the CEM.

In $\alpha \alpha$-interactions where the influence of the cascade
interactions is sufficiently large, we observe a difference of the
predicted $\pi ^-$-meson spectra. As expected, taking into account
the resonances in the FRITIOF model leads to a decreased meson yield.
The enhanced meson production in CEM becomes more pronounced in
$CC$-interactions.

To describe the proton production in the nuclear fragmentation regions
in the framework of the FRITIOF model, we take into account both
inelastic interactions of the nucleons considered above and elastic
re-scatterings. As seen, we reproduce the baryon yields in the
fragmentation regions at $y \sim 0,~~ 2.2$ for the $dd$- and $\alpha
\alpha$-interactions. The CEM calculations for $CC$-collisions were
used for determination of the nuclear destruction model parameters,
$C_{nd}=1,~~~ r_{nd} ^2=1.4$ (fm$^2$). Then we introduced the
experimental criteria for the proton registration.

The calculations for heavy nuclei look most interesting. In
Fig.  \ref{fig7} the proton distributions are presented on total and
transverse momentum in $n + Ta$ interactions at $P_n=\ 4.2$ GeV/c.
As seen, the model predictions are close to each other. The same
closeness we observe for $C + Ta$ interactions at $P =\ 4.2$
GeV/c/nucleon presented in Fig. \ref{fig8}.  Though, there is a wide
difference between the predicted meson spectra. The FRITIOF model
calculations are close to the experimental data. At the calculation we
use the following values of the parameters $C_{nd}=0.2,~~~ r_{nd}
^2=1.1$ (fm$^2$) to reproduce the $Ta$ nuclei destruction.

Summing up, we can conclude that we have reached a satisfactory
description of the meson and nucleon production in the nucleus-nucleus
interactions at the energy of 3.3 GeV/nucleon in the framework of the
sufficiently simple FRITIOF model. The model can be applied for
practical calculation of nucleus-nucleus interaction
characteristics.

The authors of the paper are thankful to Prof. V.S. Barashenkov, Zh.Zh.
Musulmanbekov and B.F. Kostenko for useful discussions. One of the
authors (V.V.U.) thanks RFBR (grand No. 00-01-00307) for its financial
support.

\newpage

\ins{fig1}{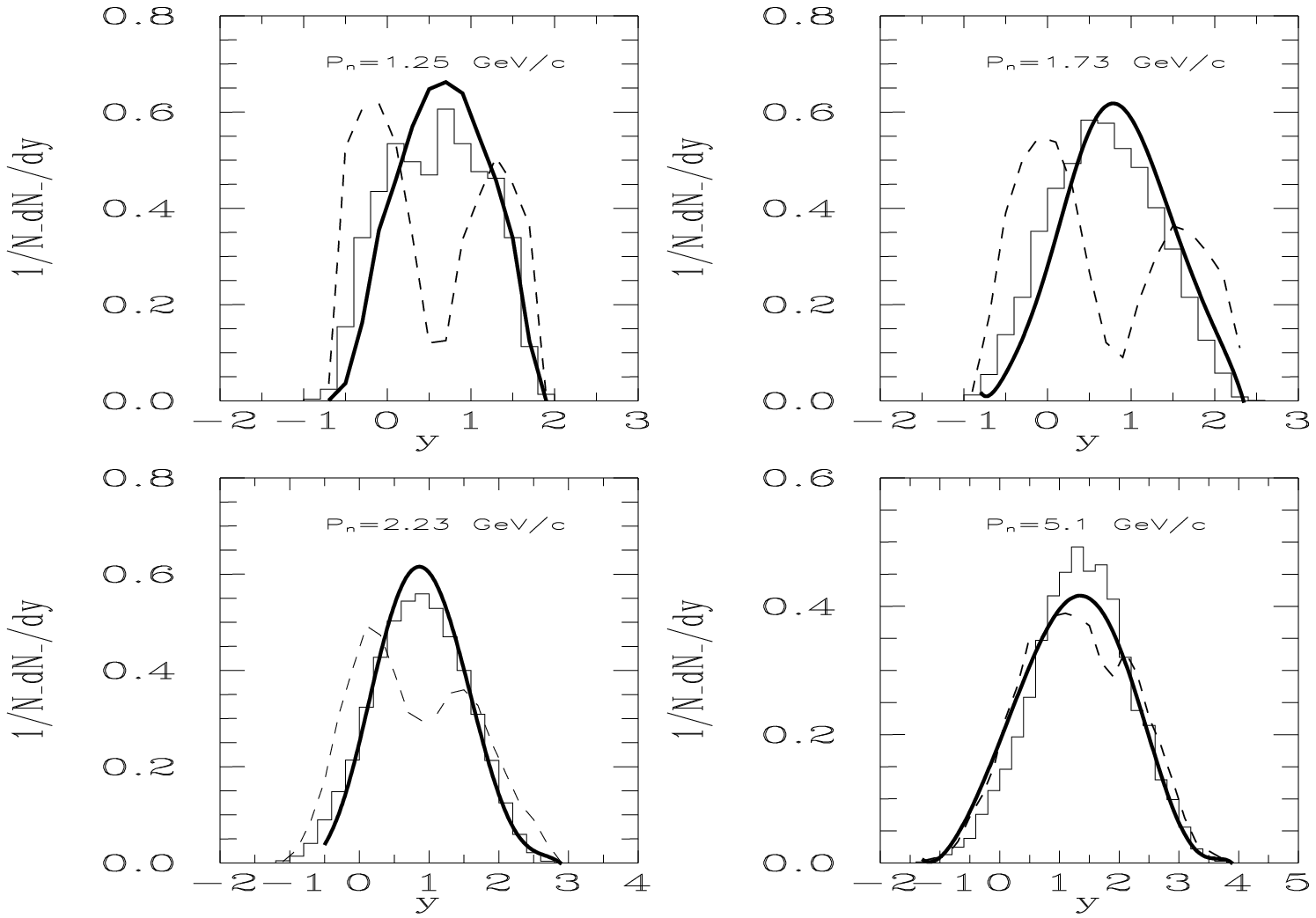}{6.1in}{4.0in}{0in}{cbth}{$\pi ^-$-meson rapidity
distributions in $np$-interactions.  Histograms are the experimental
data. The dashed and solid curves are the standard and modified FRITIOF
calculations, respectively.}

\ins{fig2}{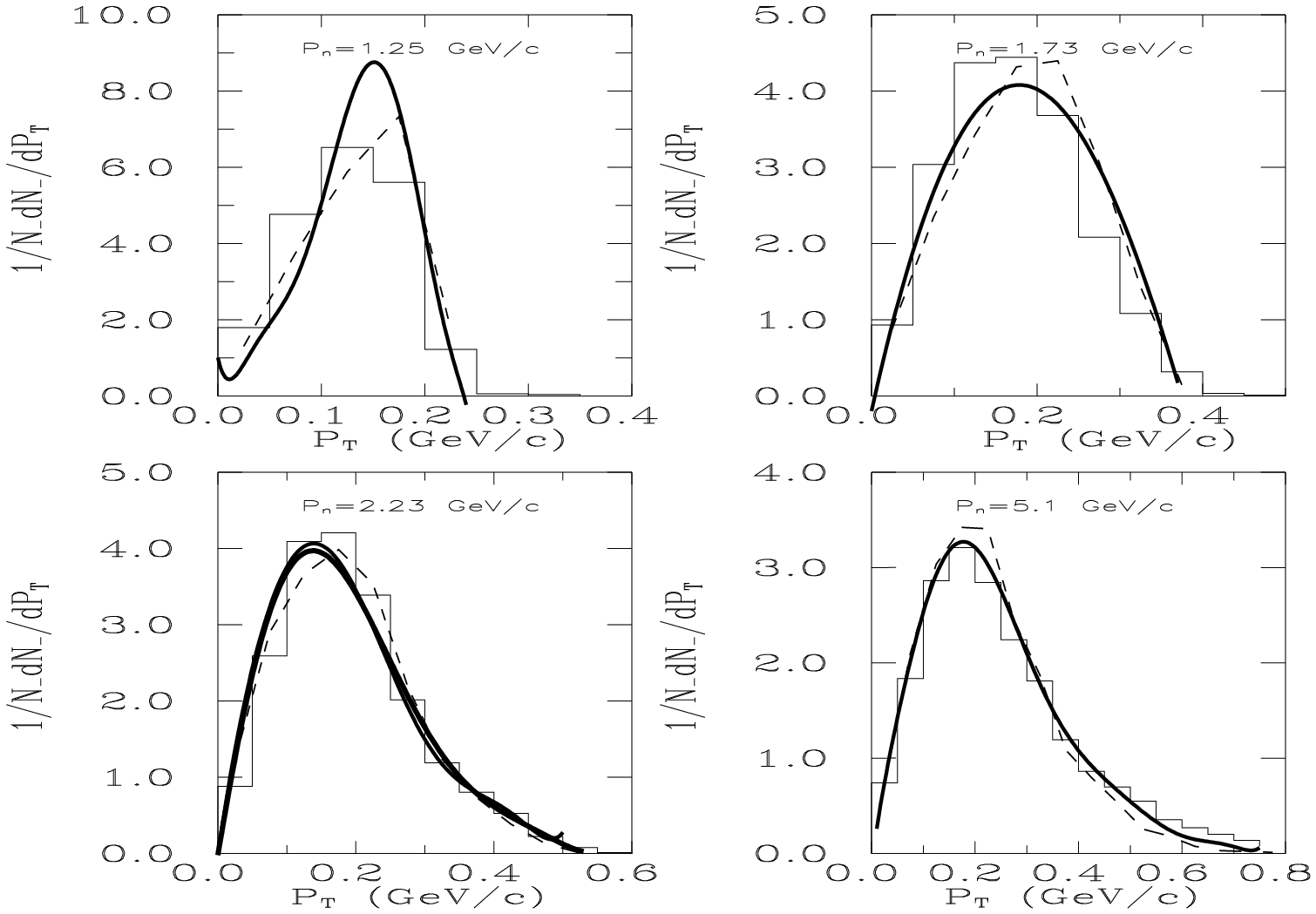}{6.1in}{4.1in}{0in}{h}{$\pi ^-$-meson transverse
momentum distributions in $np$-interactions. Notations are the same as
for Fig. 1.}

\ins{fig3}{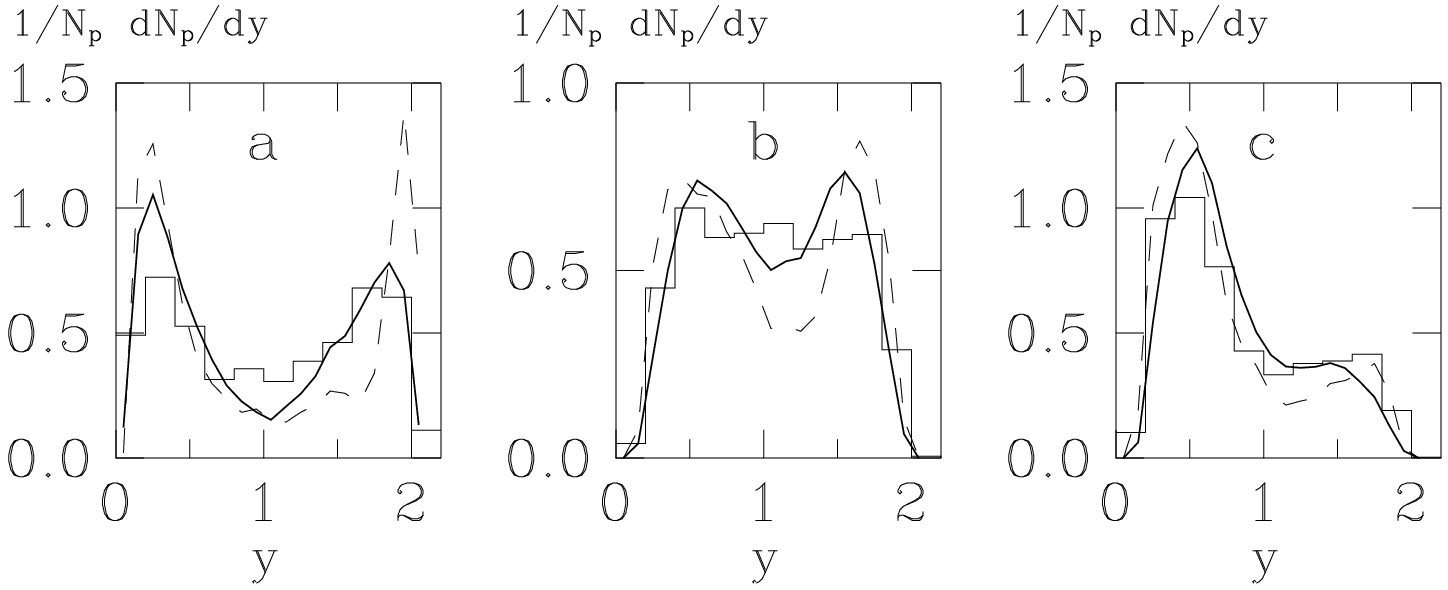}{6.1in}{2.3in}{0in}{tbch}{Proton rapidity
distributions in a) $np \longrightarrow pp\pi^-$, b) $np
\longrightarrow pp\pi^-\pi^0$, c) $np  \longrightarrow np\pi^+\pi^-$
reactions at $P_n=3.83$ GeV/c. Histograms are the experimental data.
The dashed and solid curves are the standard and modified FRITIOF
calculations, respectively.}

\ins{fig4}{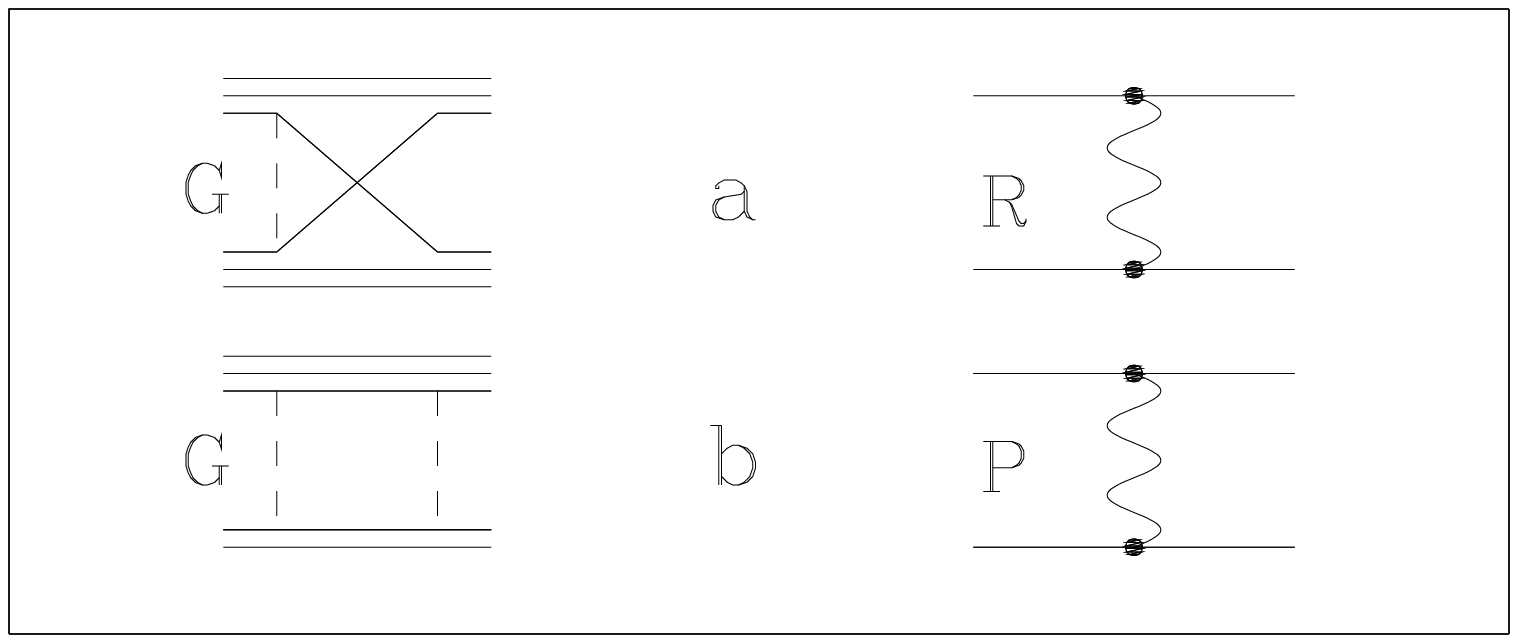}{6in}{2.5in}{0.1in}{tbch}{}
\ins{fig5}{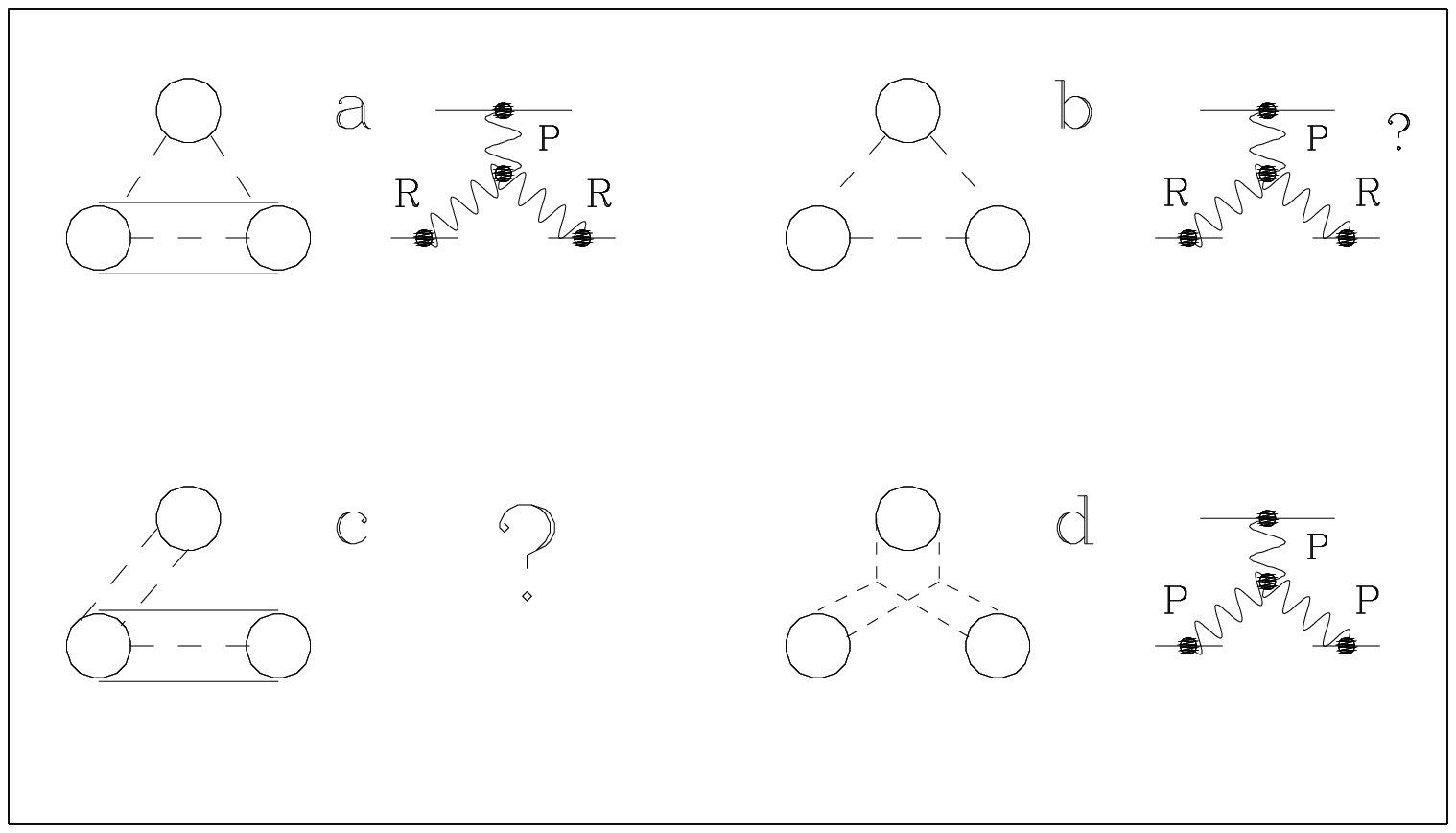}{6in}{3.5in}{0.1in}{tbch}{}

\ins{fig6}{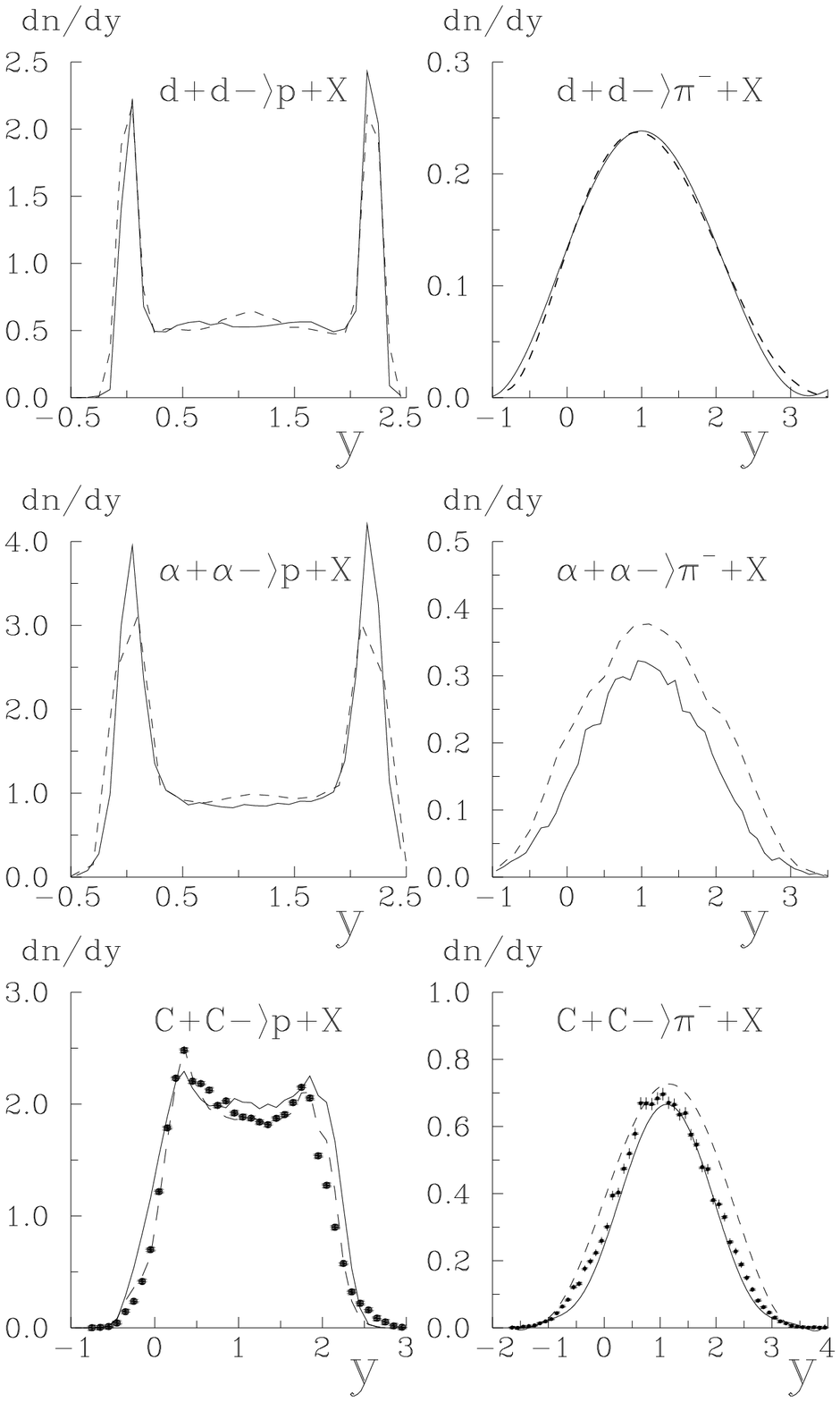}{6in}{8.8in}{0.1in}{t}{Proton and $\pi ^-$-meson
rapidity distributions in nucleus-nucleus interactions at the energy of
3.3 GeV/nucleon. Points are the experimental data [42]. The dashed and 
solid curves are CEM and the modified FRITIOF calculations, 
respectively.}

\ins{fig7}{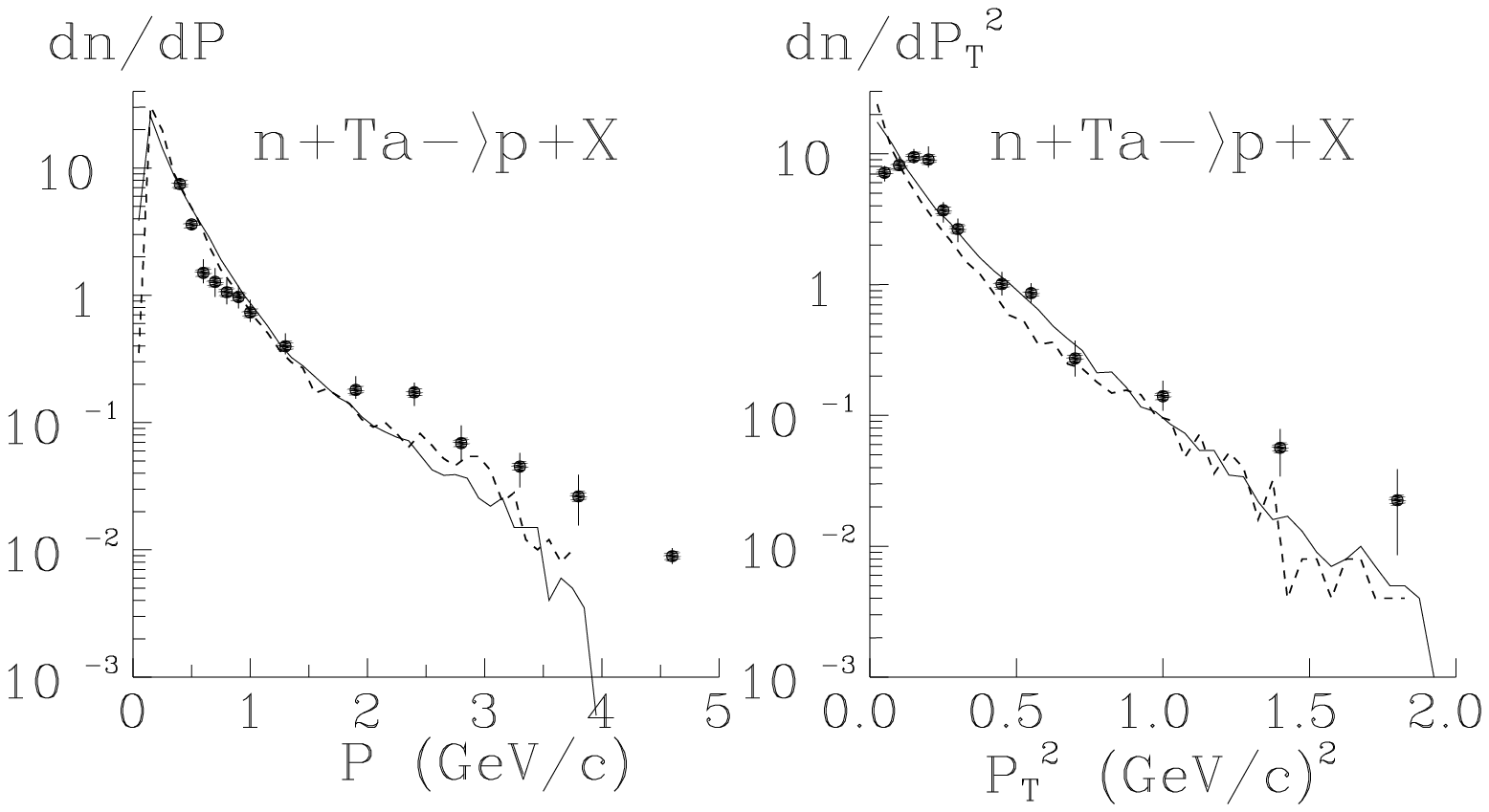}{6in}{3.2in}{0.1in}{t}{Proton rapidity
distributions in $nTa$ interactions. Points are the experimental data
[43]. The dashed and solid curves are CEM and
the modified FRITIOF calculations, respectively.}

\ins{fig8}{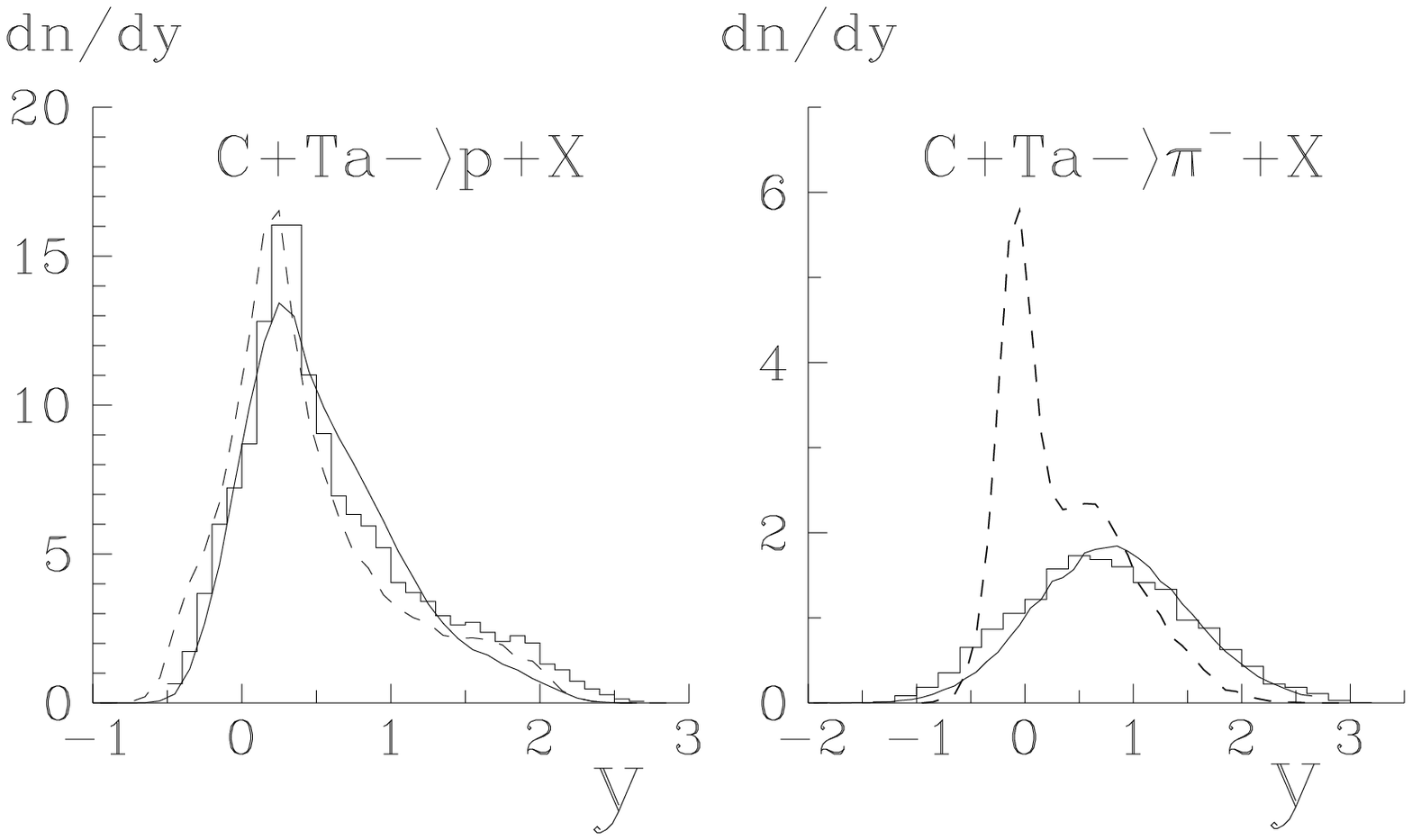}{6in}{3.5in}{0.1in}{h}{Proton and $\pi ^-$-meson
rapidity distributions in $CTa$-interactions. Points are the
experimental data [44]. The dashed and solid curves are
CEM and the modified FRITIOF calculations, respectively.}

\end{document}